\documentstyle[multicol,aps]{revtex}
\begin{document}

\title{
Spin-squeezed Ground States in the
Bilayer Quantum Hall Ferromagnet
}\par

\author{
T. Nakajima$^{1}$ and H. Aoki$^2$
}\par

\address{
$^1$Department of Physics, Tohoku University, Sendai, Miyagi 980-77,
Japan
}\par

\address{
$^2$Department of Physics, University of Tokyo, Hongo, Tokyo 113,
Japan
}\par

\date{November 5, 1997}

\maketitle

\begin{abstract}
A ``squeezed-vacuum" state considered in quantum optics
is shown to be realized in the ground-state wavefunction for
the bilayer quantum Hall system at the total Landau level filling of
$\nu=1/m$ (\,$m$:\ odd integer\,).
This is derived in the boson approximation, where a particle-hole pair
creation across the symmetric-antisymmetric gap,
$\Delta_{\rm SAS}$, is regarded as a boson.
In terms of the pseudospin describing the layers, the state
is a spin-squeezed state, where the degree of squeezing is controlled by
the layer separation and $\Delta_{\rm SAS}$.
An exciton condensation, which amounts to
a rotated spin-squeezed state, has a higher energy
due to the degraded SU(2) symmetry for $\Delta_{\rm SAS}\neq 0$.
\end{abstract}

\begin{multicols}{2}
\narrowtext

To regard a two-level system as a pseudospin has a long history.
The pseudospin ferromagnetism in the
bilayer quantum
Hall (QH) system \cite{review} is an outstanding
recent example for a bulk system.
Although the introduction of the pseudospin starts from a simple
definition of assigning the upper/lower levels (or layers)
to the pseudospin $\uparrow /\downarrow$, the ``Pauli matrices"
for them have a Lie-group structure, long known in the molecular
physics \cite{LH},
so that we have an interesting
problem of how quantum fluctuations dominate the physics in the bulk.

Here we wish to point out that
a boson ``squeezed vacuum" state considered in quantum optics
is shown to be realized in the ground-state wavefunction for
the bilayer QH system at the total Landau level filling of
$\nu=1/m$ (\,$m$:\ odd integer\,).
{\it Squeezing}, in general, is
a redistribution of quantum fluctuations between two
non-commuting observables with a preserved minimum uncertainty product.
Here the boson refers to a particle-hole pair
created across the symmetric-antisymmetric gap due to the interlayer
electron tunneling, and we can start from the vacuum of the boson.
In terms of the pseudospin describing the layers, we can also call the state
a ``spin-squeezed" state.
The squeezing is caused by the electron correlation
in the bilayer QH system, and
the degree of squeezing is controlled by two factors governing the
degradation of the pseudospin rotational symmetry,
the layer separation and $\Delta_{\rm SAS}$.

To be more precise,
the degraded rotational symmetry
results from the difference between the intralayer and interlayer
Coulomb interaction for a
finite layer separation $d$,
 and from the interlayer tunneling \cite{boson}.
Since the two layers tend to have
equal numbers of electrons
due to a capacitive charging energy,
the $z$ component of the total pseudospin
(half the difference in the numbers)
tends to vanish, $\langle s_z \rangle = 0$,
in the ground state.
Thus the pseudospin will lie in the $xy$ plane, where
the system maintains an invariance under rotations about the $z$
axis, {\it i.e.}, the SU(2) symmetry is reduced to U(1) \cite{csdl}.

The electron correlation alone pushes the ground state for
the total Landau level filling $\nu = 1/m$ (\,$m$: odd integers\,)
towards a
ferromagnetic one, {\it i.e.}, Halperin's $\Psi _{mmm}$ \cite{halperin},
so the bilayer $\nu =1/m$ QH system behaves like an easy-plane XY
itinerant-electron ferromagnet.
The U(1) symmetry is further degraded
in the presence of the interlayer tunneling,
since the tunneling amplitude
behaves like a magnetic field acting on the pseudospin,
as seen in the tunneling Hamiltonian,
$H_{\rm T}=-\Delta _{\rm SAS}\,s_x$.
This favors another pseudospin-polarized state
$\Psi _{\rm sym}$ for $\nu=1$ by pushing the electrons into the
symmetric band \cite{boson}.

So a bilayer QH system is characterized by two dimensionless parameters,
$d/\ell$ and $\Delta _{\rm SAS}/(e^2/\epsilon \ell)$ (\,$\ell$
: the magnetic length, $\epsilon$: dielectric constant\,).
The ground state for $\nu=1$ is considered to evolve continuously from
tunneling-dominated (single-particle like)
to correlation-dominated (many-body)
as $\Delta _{\rm SAS}$ is decreased,
in agreement with experimental results for
$\nu=1$ \cite{murphy}.
The fact that there is no intervening non-ferromagnetic
region between the two regimes
has been confirmed in an exact numerical calculation \cite{na1}.

Thus the energetics in the bilayer QH system
has been elaborated, which includes
the random-phase approximation (RPA) \cite{rpa}, the boson approximation
\cite{boson,na2}, macroscopic field-theoretical approaches \cite{csdl}.
However, the ground-state {\it wavefunction} has been
discussed only in the Hartree-Fock (HF)
approximation \cite{moon1}, which cannot
fully reflect the degraded symmetry of the system, and
a clear picture of wavefunctions has yet to come.
That is exactly the purpose of the present paper, where the
squeezing comes in to give the ground-state wavefunction
as a function of $d$ and $\Delta _{\rm SAS}$.
The exciton condensation, another long-standing problem in this system,
may be described in the present formalism as a
``rotated spin-squeezed" state,
which is shown to have a higher energy
due to the symmetry degraded from SU(2)
for $\Delta_{\rm SAS} \neq 0$.

We start from the effective Hamiltonian for the $\nu=1$
pseudospin ferromagnet \cite{boson,na2}.
In order to exploit the full rotational symmetry, we use a spherical
system, {\it i.e.}, an $N$-electron system on a sphere
whose surface is passed through by $2S$ flux quanta, where
$2S=N-1$ for $\nu=1$.
The Hamiltonian is given by
\begin{eqnarray}
{\cal H}_{\rm eff} &=&
\sum _{L, M} \biggl[\,(e_L+\lambda _L) \,C^{\dagger}_{L M}\,C_{LM} \nonumber \\
& & +\frac {\lambda _L}{2} \left (\,C ^{\dagger}_{L M}\,
C^{\dagger}_{L,-M} + C_{L M}\,C_{L,-M}\, \right ) \biggl] , \label{eqn:nu1} \\
e _L &\equiv & \Delta_{\rm SAS}
+\sum _{J=0}^{2S} (2J+1)\,(-1)^{2S-J}\,V_J^{\uparrow
\downarrow}\, \nonumber \\
& & \times \ 
\biggl[\,\frac {1}{2S+1}-(-1)^{2S-J}\, 
\biggl \{
\begin{array}{ccc} S & S & L\\
S & S & J
\end{array} 
\biggl \} 
\biggl ] , \\
\lambda _L &\equiv &
\sum_J^{2S-J:\,{\rm odd}} (2J+1)\,(V_J^{\uparrow \uparrow}-V_J^{\uparrow
\downarrow})\,(-1)^{2S-J}\, \nonumber \\
& & \times \ 
\biggl \{ 
\begin{array}{ccc}
S & S & L\\
S & S & J
\end{array} 
\biggl \}.
\end{eqnarray}
The interaction matrix elements can be expressed in terms of
Wigner's $6j$ symbol $\{^{S S L}_{S S J}\}$ \cite{na2}, while
$C^{\dagger}_{LM} \equiv
\sum_{j,k}\,\langle Sj;S,-k|LM\rangle \,a^{\dagger}_{j\,{\rm a}}\,
(-1)^{S-k}\,a_{k\,{\rm s}}$ creates an exciton
(a hole in the symmetric band and a particle in the
antisymmetric one)
with the total angular momentum $L$ and its $z$ component $M$,
where $\langle Sj;S,-k|LM\rangle$ is
the Clebsch-Gordan coefficient, and
$a^{\dagger}_{j\,{\rm a}}$
($a^{\dagger}_{k\,{\rm s}}$) is a creation operator for
an antisymmetric (symmetric) state
with the Landau orbit $j$\ ($k$).

The inter-particle interaction is projected onto the components
\{$V_J^{\sigma \sigma^{'}}$\} (the Haldane pseudopotential) for the
relative angular momentum $2S-J$, where $V_J^{\uparrow
\uparrow}=\,V_J^{\downarrow \downarrow}$ and $V_J^{\uparrow
\downarrow}\,=\,V_J^{\downarrow \uparrow}$ are the intra- and inter-layer
pseudopotentials, respectively.  
The terms in the form of $C^{\dagger} C$ has a coefficient
of $e_L+\lambda_L$
(involving $\Delta_{\rm SAS}$ and particle-hole correlation energies),
while we have 
off-diagonal terms (\,in the form of $C^{\dagger} C^{\dagger}$ or $C C$\,)
that create or annihilate excitons in pairs, whose coefficient
arises from the difference between the intralayer
and interlayer interactions.

In order to obtain the ground-state wavefunction, we can diagonalize
the Hamiltonian, eq.(\ref{eqn:nu1}).
Namely, we can look at how the ground state, which is
$|\Psi _{\rm sym} \rangle \equiv
\prod _j a^{\dagger}_{j\,{\rm s}}\,|0 \rangle$
in the limit of large $\Delta_{\rm SAS}$, evolves.
The starting state $\Psi _{\rm sym}$ satisfies the relation $C_{LM}\,|\Psi
_{\rm sym} \rangle = 0$ due to the definition of $C_{LM}$.
Since we have $[\,C_{LM}, C_{L'M'}\,] =
[\,C^{\dagger}_{LM},
C^{\dagger}_{L'M'}\,] = 0$,
we can call $\Psi _{\rm sym}$ the vacuum state of
the {\it boson} \{$C_{LM}$\} if
the further commutation
relation $[\,C_{LM}, C^{\dagger}_{L^{'}M^{'}}\,] =
\delta _{L\,L^{'}}\,\delta _{M\,M^{'}}$ is satisfied.
We have in fact
\begin{equation}
\left [\,C_{LM}, C^{\dagger}_{L^{'}M^{'}}\, \right ] \ \,\simeq \ \,\delta
_{L\,L^{'}}\,\delta _{M\,M^{'}} \left (\,1-\frac {N_{\rm h}}{2S+1}-\frac
{N_{\rm p}}{2S+1}\, \right ),
\end{equation}
where we have substituted
$a_{j\,{\rm s}}\,a^{\dagger}_{k\,{\rm s}}$ and
$a^{\dagger}_{j\,{\rm a}}\,a_{k\,{\rm a}}$ with their
expectation values,
$N_{\rm h}\,\delta _{j\,k}/(2S+1)$ and
$N_{\rm p}\,\delta _{j\,k}/(2S+1)$, respectively,
with $N_{\rm p}$ ($N_{\rm h}$) being the number of particles (holes).
Thus we can treat the operators \{$C_{LM}$\} as bosons
when $N_{\rm p}$
and $N_{\rm h}$ are small compared with the number of the
single-particle states (\,$=2S+1$ on a sphere\,).

The pair creation/annihilation terms in the Hamiltonian
are important in determining the electron correlation
in the ground state,
while the HF approximation neglects these terms.
Now we fully take account of these terms in the boson approximation.

A key finding here is that the diagonalization is
done with a unitary transformation, originally introduced by Bogoliubov,
{\it i.e.},
\begin{eqnarray}
{\cal H}_{\rm eff}&=&
\sum _{L,M} \omega _L\,D^{\dagger}_{LM}\,D_{LM}+E_0, \\
\omega _L &\equiv & \sqrt {e_L\,(e_L+2\lambda _L)\,}, \\
D_{LM}&\equiv &U_{\rm S}\,C_{LM}\,U^{-1}_{\rm S} \nonumber \\
&=& C_{LM}\,\cosh (\theta
_L/2)+C^{\dagger}_{L,-M}\,\sinh (\theta _L/2), \\
U_{\rm S} &\equiv& \exp \biggl[\,-\frac 14 \,\sum _{L,M} \theta _L 
\nonumber \\
& & \times 
\left
(\,C^{\dagger}_{LM}\,C^{\dagger}_{L,-M}-C_{LM}\,C_{L,-M}\,\right )
\biggl].  \label{eqn:sque}
\end{eqnarray}
Here the crucial transformation $U_{\rm S}$ is
a ``pseudo-rotation" operation belonging to
the Lie group SU(1,1) \cite{coherent},
and this indeed diagonalizes the Hamiltonian with
the angle of pseudorotation 
$\theta _L = \coth ^{-1} (1+e_L/\lambda _L)$
as far as we regard \{$C_{LM}$\} as bosons.

The ground state is then neatly expressed as
\begin{equation}
|\Psi _0 \rangle \ \,= \ \,U_{\rm S}\,|\Psi_{\rm sym} \rangle ,
\end{equation}
which involves a series of repeated creation/annihilation of exciton pairs 
if we expand the exponential form in eq.(\ref{eqn:sque}).  
We can call the state the vacuum of the transformed boson, $\{ D_{LM}\} $,
with $D_{LM}\,|\Psi _{0} \rangle =
U_{\rm S}\,C_{LM}\,|\Psi_{\rm sym} \rangle = 0$.
The energy $E_0$ appearing in eq.(5)
is the eigenenergy of $\Psi _0$.
In the language of the quantum optics,
$U_{\rm S}$ precisely corresponds to
a ``squeezing operator" \cite{umezawa}
so that we can call the ground state
a squeezed vacuum state of the original boson \{$C_{LM}$\}.

Since the boson considered here is related to
pseudospins via
the relation, $s_y \propto
\,C^{\dagger}_{00}-C_{00}$ and
$s_z \equiv \sum _j (\,a^{\dagger}_{j\,\uparrow}\,
a_{j\,\uparrow}-a^{\dagger}_{j\,\downarrow}\,a_{j\,\downarrow}\,)/2 \propto
C^{\dagger}_{00}+C_{00}$,
where $a^{\dagger}_{j\,\sigma}$
creates an electron in a $j$-th Landau orbit
with pseudospin (\,{\it i.e.}, layer index) $\sigma$,
the boson-squeezing can be recast into a pseudospin-squeezing.
For this purpose it is convenient to rotate the
pseudospin axes around the $y$ axis by $-90$ degrees
(\,$s^{'}_{x}=s_{z}$, $s^{'}_{y}=s_{y}$, $s^{'}_{z}=-s_{x}$\,),
where
$s^{'}_{z}=\sum _j (\,a^{\dagger}_{j\,{\rm a}}\,a_{j\,{\rm
a}}-a^{\dagger}_{j\,{\rm s}}\,a_{j\,{\rm s}}\,)/2$ is now half
the difference in the numbers of electrons between the antisymmetric
and symmetric states.  Then we have
\begin{eqnarray}
|\Psi _{\rm sym} \rangle &=& |\,s, s^{'}_{z}=-s \,\rangle , \\
U_{\rm S} &=& U_{\rm S}^{'}\ U(-\theta _0/2), \label{eqn:ssor} \\
U(\theta) &\equiv & \exp{\left \{\,\frac {\theta}{4s}\,
[\,(s^{'}_{+})^2-(s^{'}_{-})^2\,]\, \right \}} \nonumber \\
&=& \exp{\biggl \{\ \frac {\theta}{2} \left
[(C^{\dagger}_{00})^2-(C_{00})^2 \right ]
\biggl \}}. \label{eqn:sso}
\end{eqnarray}
Here $s$ is half the number of electrons, and
the squeezing operator $U_{\rm S}$ has been decomposed into the
$(L,M) \neq (0,0)$ component, $U_{\rm S}^{'}$,
and the $L=M=0$ component, $U(-\theta _0/2)$,
where $\theta _0 = \coth ^{-1} (1+\Delta_{\rm SAS}/\lambda _0)$ with
$\lambda _0 = \sum_J ^{\prime} (2J+1)\,
(V_J^{\uparrow \uparrow}-V_J^{\uparrow
\downarrow})/(2S+1)>0$.
The above expression, written in terms of
$s^{'}_{+}=(s^{'}_{-})^{\dagger}=\sum _j a^{\dagger}_{j\,{\rm
a}}\,a_{j\,{\rm s}}$
(the raising/lowering operators for pseudospin)
enables us to regard that
the operator $U(\theta)$
exactly corresponds to the ``spin-squeezing" operator \cite{kitaue}.

The starting state $\Psi _{\rm sym}$ has a mean pseudospin oriented
along the $s_x$ axis with circular variances of
$(\delta s_{y})^2 = (\delta s_{z})^2 = s/2$.
Thus, $\Psi _{\rm sym}$ is a spin-coherent state \cite{scs},
which is defined as a state satisfying the minimum
uncertainty relationship with variances $s/2$ equally
distributed over the two orthogonal components
normal to the mean pseudospin vector $\langle \,{\bf s}\, \rangle$ (Fig.1a).
As the ground state evolves from $\Psi _{\rm sym}$,
a spin-squeezing operator $U(\theta)$ compresses the fluctuations of
the pseudospin in one direction
at the expense of enhanced ones
in the other direction \cite{kitaue,frahm}.

We can rewrite the part,
$U(-\theta_0/2)\,|\,s, s^{'}_{z}=-s\,\rangle$, in 
a more manifestly spin-squeezed form in terms of
the spin-raising operator only \cite{frahm} as
\begin{eqnarray}
|\,\zeta \,\rangle &\equiv& U({\rm tanh}^{-1}\zeta)\,|\,s, s^{'}_{z}=-s
\,\rangle \\
&=& (\,1-|\zeta|^2\,)^{1/4}\,\exp{\biggl[\,
\frac{\zeta}{4s}\,( s^{'}_{+} )^2 \,
\biggl]}\,|\,s, s^{'}_{z}=-s \,\rangle , \label{eqn:sss1}
\end{eqnarray}
where the Campbell-Baker-Hausdorf formula is used
with $\zeta = -\tanh (\theta_0/2)$.
Then the ground state can be expressed as
\begin{equation}
|\Psi _0 \rangle = U_{\rm S}^{'}\ |\,\zeta \,\rangle . \label{eqn:zeta1}
\end{equation}

From this, we can confirm that the
spin-squeezed state $|\,\zeta\,\rangle$ with a
squeezing parameter of $\zeta=-\tanh (\theta_0/2) < 0$
has a squeezed variance
$(\delta s_{z})^2 \simeq (s/2)\,e^{-\theta_0}$ and an enhanced one
$(\delta s_{y})^2 \simeq (s/2)\,e^{\theta_0}$ for large $s$
with fixed $\langle s_z \rangle = \langle s_y \rangle = 0$.
Since $s_y$ and $s_z$ commute with
$U_{\rm S}^{'}$ within the boson approximation,
the ground state $\Psi _0$ has the same fluctuations,
$\delta s_{z}$ and $\delta s_{y}$, as a spin-squeezed
state $|\,\zeta \,\rangle$ (Fig.1b), while the
mean pseudospin is still oriented along the $s_x$ axis
for $\Delta _{\rm SAS} > 0$.

Let us now come to the physical interpretation of the pseudospin-squeezing.
In the starting state $\Psi _{\rm sym}$, pseudospin $1/2$ of
each of the $N$
electrons is oriented to the $s_x$ direction in an uncorrelated manner.
In fact, the variance $s/2 = N/4$ of each of
the two components normal to the mean
pseudospin is simply the sum of variances of
the individual pseudospins each having the variance of 1/4 \cite{kitaue}.
On the other hand, a spin-squeezing $U(-\theta _0/2)$
with $\theta _0 = \coth ^{-1} (1+\Delta_{\rm SAS}/\lambda _0)$
results from
the electron correlation ({\it i.e.},
exchange interactions and charging energies) as well as from
pair creation/annihilation of excitons across the gap $\Delta_{\rm SAS}$.
These
(nonlinear interactions in the quantum optical language)
give rise to the squeezing of the total pseudospin.
When the squeezing $U(-\theta _0/2)$ reduces
$\delta s_z$ and enhances $\delta s_y$,
the fluctuation of the azimuth angle $\delta \varphi$
of the total pseudospin
(\,the variable conjugate to $s_z$\,) is enhanced,
so that $\varphi$ tends to be ill-defined as
the squeezing becomes stronger.

We have seen that the degree of squeezing can be controlled by
varying the two parameters,
$d$ (which determines $\lambda_0$)
and $\Delta _{\rm SAS}$.
A strongly spin-squeezed state may be obtained
for a sample with
small $\Delta _{\rm SAS}$, which is easier to attain
in bilayer hole gases.
The boson approximation becomes worse as $\Delta _{\rm SAS}$
is decreased, but it has been shown that the
boson approximation gives quantitative energetics
(such as the pseudospin-wave dispersion that is either gapless or
gapful) even for small $\Delta _{\rm SAS}$ \cite{TNDron}.
We expect that the wave function is qualitatively given
by the approximation there.

If a phonon mode couples with the strongly squeezed electron state,
phonons may be concomitantly squeezed.
A phonon squeezing in a three-dimensional solid driven by light pulses
has recently been observed \cite{phonon},
but we may expect a phonon squeezing in quite a different context here.
Thus an interplay among the excitons, phonons,
and irradiated photons is an intriguing future problem.

Now we turn to the possibility of the exciton condensation
in the bilayer QH system discussed in
the literature \cite{csdl,moon1}.
In the special case of $\Delta _{\rm SAS}=0$,
the pseudospin-wave
excitation spectrum does resemble the Landau spectrum
in a superfluid $^4$He,
{\it i.e.}, having a gapless phonon-like mode
and roton-like minimum.
The exciton-condensed state may,
in the present formalism, be captured as
a ``displaced squeezed state"
\begin{eqnarray}
|\Psi _{\alpha} \rangle & \equiv &
U_{\alpha}\,|\Psi _{0} \rangle, \\
U_{\alpha} & \equiv &
\exp{\left (\,\alpha \,C^{\dagger}_{00}-
\alpha^{*}\,C_{00}\,\right )}.
\end{eqnarray}
Here $U_{\alpha}$ is a unitary displacement
operator with a complex
$\alpha$ \cite{umezawa}, again
in analogy with the quantum optics.
This state has indeed a nonzero
expectation value of an annihilation operator,
$\langle \Psi _{\alpha}|\,C_{LM}\,
|\Psi _{\alpha} \rangle = \alpha
\,\delta _{L\,0}\,\delta _{M\,0}$,
for a nonzero $\alpha$.

With this unitary transformation
the Hamiltonian reads
\begin{eqnarray}
{\cal H}_{\rm eff} &=& \sum _{L,M}
\omega _L\,D^{\alpha \,\dagger}_{LM}\,
D^{\alpha}_{LM} \nonumber \\
& & + \omega _0 \left (\,\beta ^{*}\,
D^{\alpha}_{00}+\beta \,D^{\alpha \,\dagger}_{00}\,
\right )+E_{\alpha}, \label{eqn:heff2} \\
D^{\alpha}_{LM}&\equiv &U_{\alpha}\,D_{LM}\,U^{-1}_{\alpha}\,\ =\,\
D_{LM} - \beta \,\delta _{L\,0}\,\delta _{M\,0},
\end{eqnarray}
where $\beta = \alpha\,\cosh (\theta _0/2)
+\alpha ^{*}\,\sinh (\theta _0/2)$.
We can call $\Psi _{\alpha}$ the vacuum state of
the boson \{$D^{\alpha}_{LM}$\}
since $D^{\alpha}_{LM}\,|\Psi _{\alpha}
\rangle = U_{\alpha}\,U_{\rm S}\,C_{LM}\,|\Psi _{\rm sym}
\rangle = 0$,
so that it has an energy expectation,
\begin{eqnarray}
E_{\alpha} &\equiv & \langle \Psi _{\alpha}|\,{\cal H}_{\rm
eff}\,|\Psi _{\alpha} \rangle \nonumber \\
&=& E_0+| \alpha |^2\,\left \{\,\Delta _{\rm SAS}+\lambda _0
\left [\,1+\cos (2 \phi )\,\right ]\,\right \} , \label{eqn:alpha}
\end{eqnarray}
where $\phi$ is the phase of $\alpha$
(\,$\alpha = | \alpha |\,e^{i\,\phi }$\,).
For $\Psi _{\alpha}$ to be realized
spontaneously, $E_{\alpha} \leq E_0$ is needed, but
eq.(\ref{eqn:alpha}) has the opposite
inequality since $\lambda _0 > 0$.

Then the only possibility for the exciton-condensed state to survive
is when $E_{\alpha} = E_0$.
This condition is satisfied only when $\Delta _{\rm SAS}=0$ and
$\alpha$ is pure imaginary.
When $\Delta _{\rm SAS}$ is finite,
the spontaneous symmetry breaking cannot
be expected in fact due to a finite excitation gap $\omega_0$.
The second condition that $\alpha$ be pure imaginary implies that the
displacement operator $U_{\alpha}$ amounts to a rotation about the $z$
pseudospin axis, $\exp{[\,i\,(2\,{\rm Im} \,\alpha /\sqrt {2S+1}\,)\,s_z ]}$,
 while $\beta = i\,({\rm Im}\,\alpha) \,e^{-\theta _0/2} \rightarrow 0$ and
$\theta _0 \rightarrow
+\infty$ for $\Delta _{\rm SAS} \rightarrow +0$.
Thus the exciton-condensed state corresponds,
in terms of the spin squeezing, to
a {\it rotated-squeezed} \,state,
$|\Psi (\varphi) \rangle = \exp{(-i \,\varphi s_z)}\,U_{\rm S}\,
|\Psi _{\rm sym} \rangle$,
reflecting the U(1) symmetry,
while the BCS-like state,
$\exp{(-i \,\varphi s_z)}\,|\Psi _{\rm sym} \rangle \propto \prod _j \,
( a^{\dagger}_{j\,\uparrow}+a^{\dagger}_{j\,\downarrow}\,e^{i \varphi})
|0 \rangle$, discussed in Ref.\cite{moon1} is
a {\it spin-coherent} \,state that is compatible with the SU(2) symmetry.
As seen, the exciton-condensed state
is energetically allowed only in the absence of interlayer tunneling,
where some stabilization mechanism will be further required
to make the state the true ground state.

So far we have concentrated on $\nu =1$, but
we can construct the squeezed-vacuum wavefunction
for fractional values of Landau level filling
$\nu=1/m$ (\,$m=3,5,\cdots$\,)
via a composite-fermion transformation \cite{naPRL} 
as has been proposed for
bilayer systems \cite{na3} by the present authors.
In a spherical system this can readily be done, since 
everything is written in terms of 
angular momenta.

We are grateful to Shin Takagi, Tetsuo Ogawa, Satoru Okumura and Yasuo
Okajima for illuminating discussions on the coherent state,
and to Komajiro Niizeki for valuable comments.
This work was in part supported by a Grant-in-Aid from the Ministry of
Education, Science and Culture, Japan.

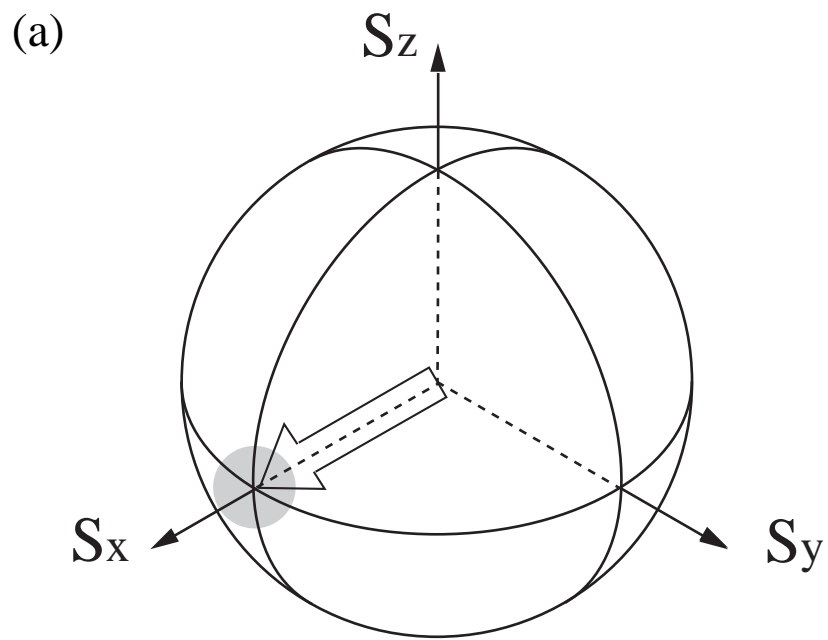
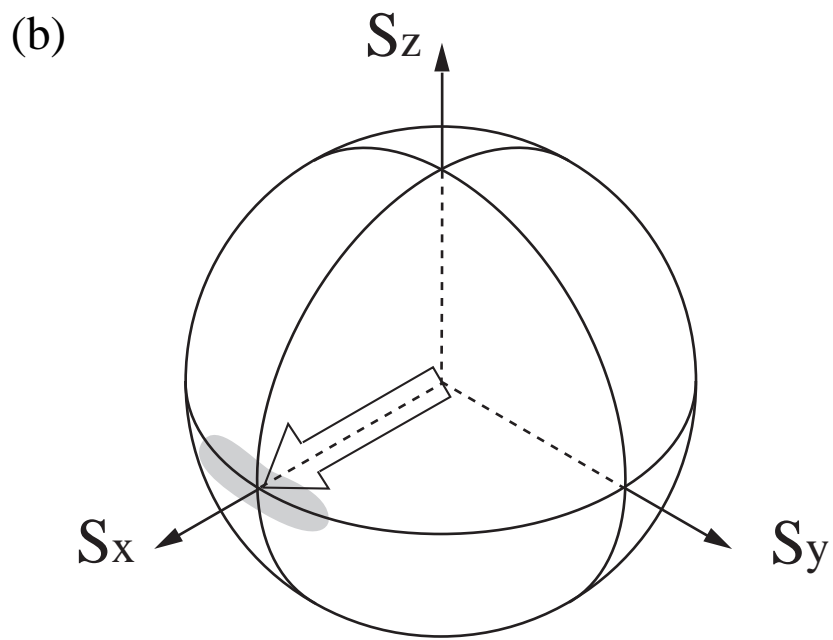
\begin{figure}
\caption{(a) A spin-coherent state (corresponding
to $\Psi_{\rm sym}$)
and (b) a spin-squeezed state ($\Psi_0$)
are schematically shown, where
the regions over which spins fluctuate are indicated by shading.}
\end{figure}

\end{multicols}

\end{document}